\documentclass[12pt]{article}
\usepackage{latexsym}
\usepackage{amsmath,,calrsfs}
\usepackage{amsfonts}
\usepackage{amssymb}
\usepackage{amscd}
\usepackage{fancybox}
\usepackage{cite}
\usepackage{amsmath,amsfonts,amsbsy}
\usepackage{pstricks,pst-node}
\usepackage[small,bf,hang]{caption2}
\usepackage{psfrag}

\usepackage{float}

\psset{unit=1.3cm,linewidth=.5pt,radius=.2}  

\usepackage{enumitem}                       
\usepackage{multirow}                     
\usepackage{float}                          
\usepackage{lscape}                         
\usepackage{bm}

\def\hybrid{\topmargin -20pt    \oddsidemargin 0pt
        \headheight 0pt \headsep 0pt
        \textwidth 6.25in       
        \textheight 9.5in       
        \marginparwidth .875in
        \parskip 5pt plus 1pt   \jot = 1.5ex}

\hybrid

\newcommand{\bb}{\bar\beta}

\newcommand{\beq}{\begin{equation}}
\newcommand{\eeq}{\end{equation}}
\newcommand{\bi}{\begin{itemize}}
\newcommand{\ei}{\end{itemize}}
\newcommand{\bt}{\begin{tabular}}
\newcommand{\et}{\end{tabular}}
\newcommand{\bc}{\begin{center}}
\newcommand{\ec}{\end{center}}

\newcommand{\be}{\begin{equation}}
\newcommand{\ee}{\end{equation}}
\newcommand{\bea}{\begin{eqnarray}}
\newcommand{\eea}{\end{eqnarray}}
\newcommand{\ba}{\begin{array}}
\newcommand{\ea}{\end{array}}

\def\bbox{{\,\lower0.9pt\vbox{\hrule \hbox{\vrule height 0.2 cm
\hskip 0.2 cm \vrule height 0.2 cm}\hrule}\,}}
\newcommand{\dsl}{\pa \kern-0.5em /}

\font\mybb=msbm10 at 12pt
\def\bb#1{\hbox{\mybb#1}}

\def\bP {\bb{P}}
\def\bQ {\bb{Q}}

\begin{document}

\begin{titlepage}
\begin{center}

\hfill  DAMTP-2012-49\\
\hfill LMU-ASC 49/12

\vskip 1.5cm

{\Large \bf On the Hamiltonian form  of 3D massive gravity}

\vskip 1cm

{\bf Olaf Hohm$^1$, Alasdair Routh$^2$, Paul K. Townsend$^2$ \\ 
and Baocheng Zhang$^{2,3}$}

\vskip 25pt

{\em $^1$ \hskip -.1truecm Arnold-Sommerfeld-Center for Theoretical Physics, 
Fakult\"at f\"ur Physik, Ludwig-Maximilians-Universit\"at M\"unchen, 
Theresienstrasse 37, 80333 M\"unchen, Germany
\vskip 5pt }

{email: {\tt olaf.hohm@physik.uni-muenchen.de}} \\

\vskip 15pt

{\em $^2$ \hskip -.1truecm Department of Applied Mathematics and Theoretical Physics,\\ Centre for Mathematical Sciences, University of Cambridge,\\
Wilberforce Road, Cambridge, CB3 0WA, U.K. \vskip 5pt }

{email: {\tt A.J.Routh@damtp.cam.ac.uk, P.K.Townsend@damtp.cam.ac.uk}} \\

\vskip 15pt

{\em $^3$ \hskip -.1truecm State Key Laboratory of Magnetic Resonances and Atomic and
Molecular Physics, \\ Wuhan Institute of Physics and Mathematics,\\ 
	Chinese Academy of Sciences, \\
Wuhan 430071, China\\ 
	 \vskip 5pt }

{e-mail: {\tt zhangbc@wipm.ac.cn}}

\end{center}

\vskip 0.5cm

\begin{center} {\bf ABSTRACT}\\[3ex]\end{center}

We present a ``Chern-Simons-like''  action for the ``general massive gravity'' model propagating 
two spin-2 modes with independent masses in three spacetime dimensions (3D), and we use it to
find a simple Hamiltonian form of  this model. The number of local degrees of freedom,  determined by the 
dimension of the physical phase space, agrees with a linearized analysis except in some limits,   in particular
that yielding ``topologically new massive gravity'',  which therefore suffers  from a  linearization instability. 

\end{titlepage}

\newpage
%

\section{Introduction}
\setcounter{equation}{0}

Massive gravity models have been intensively investigated over the past few years, 
partly motivated by the idea that some 
infra-red modification of  General Relativity (GR) could provide an alternative to dark energy; see e.g. \cite{Hinterbichler:2011tt} for an overview.  As for many other issues in GR, it is useful to consider how things simplify in the context of a three-dimensional (3D) spacetime. 
Although the long-standing  ``topologically massive gravity''  (TMG) model \cite{Deser:1981wh} shows that a massive  graviton is possible in 3D,   this is achieved by the introduction of the  parity violating,  Lorentz-Chern-Simons  (LCS)  term, which is the action for 3D conformal gravity \cite{vanNieuwenhuizen:1985cx}. In contrast, the more recent  ``new massive gravity'' (NMG) model \cite{Bergshoeff:2009hq,Bergshoeff:2009aq}  achieves a similar effect without breaking parity through the introduction of a particular  curvature-squared term;  this model exhibits in simplified form some of the features of ghost-free higher-dimensional models of massive gravity (see e.g.  \cite{deRham:2011ca,Paulos:2012xe,deRham:2012az} for a discussion of this point).   The inclusion of both the LCS term  and the  curvature-squared term of NMG leads to the ``general massive gravity'' (GMG) model \cite{Bergshoeff:2009hq}.  

Allowing for a cosmological term, and omitting an overall  factor proportional to the inverse 3D Newton constant,  the Lagrangian density for GMG, with spacetime metric $g_{\mu\nu}$ ($\mu=0,1,2$),  takes the form
\be
{\cal L}= \sqrt{|g|} \left[ \sigma R - 2 \Lambda_0 +  \frac{1}{m^2} G^{\mu\nu}S_{\mu\nu}\right] + \frac{1}{\mu} {\cal L}_{LCS}\, , 
\ee
where the tensors $G$ and $S$ are, respectively, the Einstein and (3D) Schouten tensors,  $m$ and $\mu$ are two mass parameters, $\sigma$ is a dimensionless constant and $\Lambda_0$ is the cosmological parameter; in a maximally symmetric vacuum we have $G_{\mu\nu} = -\Lambda g_{\mu\nu}$, where the cosmological constant $\Lambda$ is given by
\be\label{cosconstant}
\Lambda = -2m^2\left[\sigma  \pm \sqrt{\sigma^2+ \lambda}\right]\, , \qquad \lambda= \Lambda_0/m^2\, . 
\ee
Notice that  $\Lambda_0=0$ allows a Minkowski vacuum, in which case perturbative unitarity requires $\sigma\le0$; in other  vacua it is useful to allow for  arbitrary $\sigma$ (although for $\sigma\ne0$ one may rescale fields such that  $\sigma^2=1$). 
Linearization about the Minkowski vacuum leads to a generalization of the 3D Fierz-Pauli (FP) equations that propagates  two spin-2 modes of  independent masses;  NMG corresponds to the equal mass case, and TMG to the case in which  one mode is decoupled by taking its mass to infinity.  Linearization about other vacua leads to modified versions of  these equations that depend on the ratios of the masses $(\mu,m)$  to the scale set by the cosmological constant.  

While a linear approximation usually allows a reliable count of the number of local degrees of freedom, there are cases in which it gives misleading results. A Hamiltonian formulation provides a  way to count the number of local degrees of freedom without resort to linearization:  this number may be defined as half the dimension per space point  of the physical phase space (i.e.~taking into account all constraints and gauge invariances).  For example,  using the standard 
Arnowitt-Deser-Misner (ADM) Hamiltonian formalism for GR in a spacetime of dimension $D$, this definition tells us that there are  $D(D-3)/2$ degrees of freedom, which coincides with the number of polarization states of a massless graviton.
An extension of the ADM formalism to higher-derivative gravity theories in any spacetime dimension was worked out in \cite{Deruelle:2009zk} but it is not applicable to parity-odd 3D theories like TMG and GMG.   The Hamiltonian formulation of TMG has been studied in many papers  e.g.  \cite{Deser:1991qk,Buchbinder:1992pe,Grumiller:2008pr} but of most relevance here is the formulation that we shall refer to as the ``minimal''  formulation of  Carlip \cite{Carlip:2008qh}, some aspects of 
which were clarified  by Blagojevic and Cvetkovic  \cite{Blagojevic:2008bn} via an implementation of  Dirac's general procedure  \cite{Dirac:1950pj}. 

More recently, Dirac's  procedure was used to find a Hamiltonian formulation of NMG  \cite{Blagojevic:2010ir}, which was subsequently 
applied to the case in which $\lambda=-\sigma^2$ \cite{Blagojevic:2011qc}. This case is special for many reasons \cite{Bergshoeff:2009aq} 
but the one of relevance here is that linearization yields equations with an accidental gauge invariance that describes a  ``partially massless'' graviton  with only one local degree of freedom \cite{Deser:1983mm,Deser:2001pe}. The gauge invariance is a linearized Weyl invariance, which is ``accidental'' in the sense that it does not extend beyond the linearized approximation, and for this 
reason one should suspect the reduction in the number of local degrees of freedom to be an artefact of the linear approximation. Indeed, it turns out that there is no reduction 
in the number of local degrees of freedom if this number is counted using the Hamiltonian formulation of NMG. This implies that NMG suffers from a linearization instability
at ``partially massless vacua''. This result has been extended to GMG in \cite{Afshar:2011qw}. 
 
There is one other case in which linearization leads to an accidental  gauge invariance, which is again a linearized Weyl invariance. This is the case in which 
$\Lambda_0=0$ and $\sigma=0$;  the linearized Weyl invariace  is ``accidental''  because the non-linear theory is definitely not Weyl invariant \cite{Bergshoeff:2009hq}.  The special subcase in which $\mu=\infty$ was first analysed by Deser \cite{Deser:2009hb}, who showed that it propagates a single massless mode; we shall call this model ``massless NMG''.  The generic case was  analysed in \cite{Andringa:2009yc}, where it was called ``topologically new massive gravity'' (TNMG); this model was studied independently (under a different name)  in \cite{Dalmazi:2009pm}. Linearized TNMG propagates a single spin-2 mode 
of mass  $m^2/\mu$, which  becomes the massless mode of ``massless NMG'' in the $\mu\to\infty$ limit. There is therefore an apparent reduction in the number of local degrees of freedom\footnote{The natural extension of TNMG  to adS vacua is ``chiral GMG'', which occurs for $\lambda= -3\sigma^2 + 2\rho\left(\rho-3\sigma\right) -2\left(\rho -2\sigma\right) \sqrt{\left(\rho-2\sigma\right)\rho}$, where $\rho= m^2/\mu^2$; this is the GMG analog of the perturbative unitary limit $\lambda=-3\sigma^2$ of  NMG \cite{Bergshoeff:2009aq}.   Although there are discontinuities in the spectrum of linearized  GMG  at the chiral point, 
the  linearized approximation does not lead to any reduction in the {\it number} of local degrees of freedom, except in the Minkowski limit where it degenerates to TNMG.} 
but,  as in the ``partially massless'' case, we should expect this reduction to be an artefact of the linearized approximation.  One of the aims of  this paper is to use a Hamiltonian formulation to verify this. 

Implementation of Dirac's procedure for 3D massive gravity models leads to a rather large phase space with a correspondingly large number of constraints of both ``first-class'' and ``second-class''  in  Dirac's terminology (first-class constraints being those that generate gauge transformations).  Another purpose of this paper is to present a simple Hamiltonian formalism of NMG and GMG that, like Carlip's Hamiltonian formulation of TMG  \cite{Carlip:2008qh},  is ``minimal'' in the sense that  (i) the only first-class constraints are the six that generate 3D diffeomorphisms and local Lorentz transformations,  and (ii) the  number of second-class constraints is minimized.

Our starting point  is a new action for  GMG, and its TMG limit, that  is  ``Chern-Simons-like'' in the sense that it is  the integral of a Lagrangian 3-form constructed by taking exterior products of independent one-form fields and their exterior derivatives, and so does not require a metric for its construction.  As in Chern-Simons (CS) gravity models, the one-form fields include the  Lorentz frame, or dreibein, one-forms $e^a$ ($a=0,1,2$) which can be used to construct a metric if we assume that its matrix of coefficient functions $e_\mu{}^a$ 
is invertible. In a decoupling limit in which  all propagating modes become infinitely massive, the action reduces to the Einstein-Cartan (EC) formulation of 3D GR, which is a CS gravity model \cite{Achucarro:1987vz,Witten:1988hc}.  In the partial decoupling limit that yields TMG, the action reduces to the sum of the EC action and the CS action for 3D conformal gravity of Horne and Witten \cite{Horne:1988jf}. 

It might seem superfluous to say that the Hamiltonian formulation of massive gravity models breaks the manifest  spacetime diffeomorphism invariance of the covariant action because the Hamiltonian formulation involves a distinction between time and space. However, the Hamiltonian form of the action for a CS theory is just the covariant action rewritten in a non-covariant way
by performing a time/space decomposition of all fields. In contrast, the Hamiltonian form of the action  of a  CS-like massive gravity model cannot be obtained in this way; there are necessarily additional, secondary-constraint, terms that break this manifest spacetime covariance. While one such constraint is required for TMG \cite{Carlip:2008qh}, two are required for GMG. 

Using our  Hamiltonian formulation of GMG,  we are able to compute the dimension of the physical phase space and to see how this depends on the parameters 
of the model. In particular, we are able to show that the number of local degrees of freedom changes only in a decoupling limit and is therefore not discontinuous in any of the special limits discussed above, except in the limit that yields conformal 3D gravity. This means, in particular, that both the ``massless NMG'' limit of NMG and the 
TNMG limit of GMG suffer from linearization instabilities.

\section{Preliminaries}
\setcounter{equation}{0}

As our covariant starting point for a Hamiltonian formulation of massive 3D gravity models is a Chern-Simons-like action 
for these models, we review here the aspects of the Chern-Simons gravity models that will be of relevance to us. This will also serve
to introduce our conventions and terminology. 

\subsection{Einstein-Cartan} 

In the Einstein-Cartan (EC) formulation of 3D gravity, the  independent fields are the dreibein $e^a$ ($a=0,1,2$), a Lorentz-vector valued one-form, and an adjoint-valued local Lorentz connection one-form which we may trade for the vector-valued one-form  $\omega^a$. From these one may construct the Lorentz-covariant torsion and curvature two-forms
\be
T^a = D e^a \equiv de^a + \epsilon^{abc} \omega_b e_c\, , \qquad  
R^a = d\omega^a  + \frac{1}{2} \epsilon^{abc} \omega_b\omega_c\, , 
\ee
where $\epsilon^{abc}$ is the antisymmetric invariant tensor of $SO(1,2)$, with $\epsilon^{012}=1$. Here,  and henceforth,  products of forms should be understood as exterior products. The torsion and curvature two-forms satisfy the Bianchi identities
\be
DT^a \equiv \epsilon^{abc}R_be_c \, , \qquad DR^a \equiv 0\, . 
\ee
This formalism  allows the 3D Einstein-Hilbert action, with cosmological term, to be written in a first-order form as an integral of the Einstein-Cartan (EC)  Lagrangian  3-form
\be\label{EC3form}
L_{EC} = -\sigma e_aR^a + \frac{\Lambda_0}{6} \epsilon^{abc} e_a e_b e_c\, . 
\ee
As explained in the introduction, the constant $\sigma$ is included for later convenience; $\sigma=1$ is the standard choice  in 3D GR  for 
our ``mostly plus'' metric  signature convention. The EC action is already first-order in time derivatives and so constitutes a simple starting point for the Hamiltonian formulation of GR \cite{Deser:1976ay}. 

The field equation for  $\omega^a$  is $T^a=0$. If we assume that the dreibein matrix  $e_\mu{}^a$  is invertible, with inverse $e_a{}^\mu$,  then we can solve for 
$\omega^a$:
\be
\omega_\mu{}^a  =  - e^{-1} \varepsilon^{\nu\rho\sigma}\left(e_\nu{}^a e_{\mu\, b} - \frac{1}{2} e_\mu{}^a e_{\nu\, b} \right)\partial_\rho e_\sigma{}^b \equiv \omega_\mu{}^a (e)\, , 
\qquad (e= \det e_\mu{}^a), 
\ee
where $\varepsilon^{\mu\nu\rho}$ is the totally antisymmetric  invariant tensor density. Let us note here for future use that
\be
e^{-1} \varepsilon^{\mu\nu\rho} = \epsilon^{abc} e_a{}^\mu e_b{}^\nu e_c{}^\rho \equiv \epsilon^{\mu\nu\rho}\, . 
\ee
Under the same assumption of invertible  dreibein, we have that 
\be\label{identity}
\frac{1}{2}\varepsilon^{\mu\nu\rho} R_{\nu\rho}{}^a = -e\,  e_\nu{}^a G^{\nu\mu}\, ,
\ee
where $G_{\mu\nu}$ is the Einstein tensor constructed from the metric  $g_{\mu\nu} := e_\mu\cdot e_\nu \equiv e_\mu{}^a e_\nu{}^b \eta_{ab}$  and the affine connection
\be
\Gamma_{\mu\nu}{}^\lambda = \omega_{\mu\, a}{}^b e_\nu{}^a e_b{}^\lambda + e_a{}^\lambda \partial_\mu e_\nu{}^a\, ,  
\ee
which becomes the standard Levi-Civita connection for the metric $g_{\mu\nu}$ when $\omega=\omega(e)$.  Using the fact that $2g^{\mu\nu}G_{\mu\nu} =-R$ in 3D, where $R$ is the Ricci scalar, we see that elimination 
of $\omega^a$ yields the  Lagrangian density
\be
{\cal L} = \frac{1}{2}e\left[\sigma R - 2\Lambda_0\right]\, . 
\ee
The identity  (\ref{identity}) may also be used to show that  the $e^a$-equation is, given $\omega^a=\omega^a(e)$,  the Einstein equation $\sigma G_{\mu\nu} = -\Lambda_0 g_{\mu\nu}$, from which we see that $\Lambda_0/\sigma$ is the cosmological constant for the EC model . 

A variant of the EC action, which will be useful for later purposes, is defined by the Lagrangian 3-form 
\be\label{variantEC}
\tilde L_{EC} = L_{EC} + h_a T^a\, , 
\ee
where the new Lorentz-vector-valued one-form  $h^a$ is a Lagrange multiplier for the constraint $T^a=0$. The equivalence of this action to the standard one can 
be seen by rewriting it in terms of $e^a$ and  $\Omega^a = \omega^a - h^a/\sigma$; one finds that 
\be
\tilde L_{EC} = L_{EC}(e,\Omega) + \frac{1}{2\sigma} \epsilon^{abc} e_a h_b h_c\, . 
\ee
In this action the field $h^a$  can be trivially eliminated (given  invertibility of the dreibein) whereupon the action 
reduces to the standard EC action.

\subsection{Horne-Witten}

Now consider the Lagrangian 3-form
\be\label{HW}
L_{HW} =  \frac{1}{2} \omega_a d\omega^a + \frac{1}{6} \epsilon_{abc} \omega^a\omega^b\omega^c + h_a \left(T^a -b e^a\right)
+\frac{1}{2} bdb \, .
\ee
This is the  CS theory for the 3D conformal group constructed by Horne and Witten  \cite{Horne:1988jf}.  In addition to the one-form gauge potentials $(e^a,\omega^a)$ of the EC model we have  an additional Lorentz-vector valued one-form potential $h^a$ associated to proper conformal gauge transformations, and a Lorentz-scalar one-form potential $b$ associated to local scale (Weyl) transformations. The relative coefficients are fixed, up to field redefinitions, by the requirement  of local $SO(2,2)$ gauge invariance.

The $(h^a,\omega^a)$ fields are auxiliary fields in the HW model in the sense that they may be eliminated by their  field equations, which are jointly  equivalent to 
\be\label{twoeqs}
T^a=be^a\, , \qquad R^a + \epsilon^{abc} e_b h_c =0\, . 
\ee
Assuming invertibility of $e_\mu{}^a$,  these equations imply that
\be\label{homega}
\omega_\mu{}^a = \omega_\mu{}^a (e,b) \equiv   \omega_\mu{}^a(e) + \epsilon_\mu{}^{a\nu}b_\nu\, , \qquad 
h_{\mu\nu} =  -S_{\mu\nu}(e,b)\, , 
\ee
where $S_{\mu\nu}(e,b)$ is the Schouten tensor for the connection $\omega_\mu{}^a(e,b)$; recall that, in 3D, 
\be
S_{\mu\nu}=  R_{\mu\nu} - \frac{1}{4} g_{\mu\nu} R\, , 
\ee
where $R_{\mu\nu}$ is the Ricci tensor and  $R$ the Ricci scalar. 
Substituting for $\omega$ using the first of equations (\ref{homega}), one finds that 
\begin{eqnarray}
S_{\mu\nu}(e,b) &=&  S_{\mu\nu}(e) + \bar D_\mu b_\nu + \left(b_\mu b_\nu - \frac{1}{2}g_{\mu\nu} b^2\right)\, ,  
\end{eqnarray}
where $S_{\mu\nu}(e)$ is the (symmetric) Schouten tensor for the standard torsion-free connection, and  $\bar D$ indicates a covariant derivative with respect to this  connection. 
Observe that 
\be
S_{[\mu\nu]}(e,b) = \partial_{[\mu}b_{\nu]}\, , 
\ee
from which we see that  the second of equations (\ref{homega}) implies that
\be\label{beq}
h_{[\mu\nu]} = -\partial_{[\mu} b_{\nu]}\, . 
\ee
This is  also the $b$-equation of motion. Finally, the $e^a$-equation is 
\be\label{HWeq}
\left(D+b\right)h^a=0\, . 
\ee
Upon substituting the expressions for $(h^a,\omega^a)$, one finds that  all $b$-dependence cancels and we are left with the equation  $C_{\mu\nu}=0$, where the left hand side is the symmetric and traceless Cotton tensor
\be
C_{\mu\nu}=  \epsilon_\mu{}^{\tau\rho} \bar D_\tau S_{\rho\nu}(e) \, . 
\ee
This  is the standard field equation of 3D conformal gravity. Off-shell equivalence of the HW model to conformal gravity can also be shown by using (\ref{homega}) to eliminate $(h^a,\omega^a)$ from the HW action.  One finds that the $b$ field drops out and the resulting action is  the Lorentz-Chern-Simons term for the composite connection $\omega(e)$.  


\section{Massive 3D gravity} 
\setcounter{equation}{0}

Consider the following CS-like Lagrangian 3-form 
\begin{eqnarray}\label{LagGMG2}
L_{GMG} &=& -\sigma e_a R^a  + \frac{1}{6} \Lambda_0 \epsilon^{abc} e_a e_b e_c 
+ h_a T^a  + \frac{1}{2\mu}\left[ \omega_a d\omega^a  + \frac{1}{3} \epsilon^{abc}\omega_a\omega_b\omega_c\right]\nonumber \\
&&\  - \ \frac{1}{m^2} \left[ f_a R^a + \frac{1}{2}\epsilon^{abc} e_a f_b f_c \right] \, . 
\end{eqnarray}
In the limit  in  which both $\mu\to\infty$ and $m\to\infty$ we get the variant EC Lagrangian 3-form of  (\ref{variantEC}).
In the limit that $m\to\infty$ for finite $\mu$ we get the Lagrangian 3-form
\be\label{LagTMGs}
L_{TMG} = -\sigma e_a R^a  + \frac{1}{6} \Lambda_0 \epsilon^{abc} e_a e_b e_c 
+ h_a T^a  + \frac{1}{2\mu}\left[ \omega_a d\omega^a  + \frac{1}{3} \epsilon^{abc}\omega_a\omega_b\omega_c\right]\, , 
\ee
which  is known to describe TMG \cite{Grumiller:2008pr,Carlip:2008qh}.  

In the generic case of finite $\mu$ and finite $m$ the above Lagrangian 3-form is equivalent to the one proposed for GMG
in \cite{Blagojevic:2010ir,Afshar:2011qw} if one assumes an invertible dreibein field, but it  seems not to have been previously appreciated that the GMG Lagrangian can be written in the above way.  Let us first confirm that it  does indeed describe GMG.  The $(h^a,\omega^a,f^a)$ equations are jointly equivalent to
\be
T^a=0\, , \qquad - \frac{1}{m^2} Df^a + \frac{1}{\mu} R^a +  \epsilon^{abc} e_b h_c =0 \, , \qquad 
R^a + \epsilon^{abc} e_b f_c =0\, , 
\ee
which imply that
\be\label{three}
\omega_\mu{}^a= \omega_\mu{}^a(e) \, , \qquad f_{\mu\nu}= -S_{\mu\nu}(e)\, , \qquad h_{\mu\nu} = -\frac{1}{\mu} S_{\mu\nu}(e)-\frac{1}{m^2} C_{\mu\nu}(e)\, , 
\ee
and hence that 
\be\label{secondary}
h_{[\mu\nu]} =0\, , \qquad f_{[\mu\nu]}=0\, . 
\ee
Back substitution yields the  Lagrangian density
\be
{\cal L} _{GMG} = \frac{1}{2}e\left\{ \sigma R - 2\Lambda_0 +\frac{1}{m^2}  G^{\mu\nu}(e) S_{\mu\nu}(e)\right\} 
+ \frac{1}{\mu} {\cal L}_{LCS}\, , 
\ee
where ${\cal L}_{LCS}$ is the standard Lorentz-Chern-Simons term of 3D conformal gravity. For  $\sigma=-1$ this defines GMG, and NMG is obtained in the $\mu\to\infty$ limit. 

\subsection{Hamiltonian preliminaries}

The action defined by integration of  (\ref{LagGMG2}) over a 3-manifold with a Cauchy 2-surface is a  starting point for a Hamiltonian formulation of GMG, and its various limits.  Observe that  the Lagrangian density corresponding to 
the Lagrangian 3-form (\ref{LagGMG2})  takes the general form
\be\label{TopLag}
{\cal L}= \frac{1}{2}  g_{rs} a^r \cdot d a^s + \frac{1}{6}f_{rst} \, a^r \cdot a^s \times a^t\, , 
\ee
where $a^r$ ($r=1,\dots,N$) are $N$ `flavours' of 3-vector one-forms, and we use  a 3D Lorentz-vector algebra notation
in which, e.g. 
\be
a^p \cdot a^q = \eta^{ab} a_a^p a_b^q\, , \qquad \left(a^s \times a^t \right)^a = \epsilon^{abc} a_b^s a_c^t\, , 
\ee
where $\eta$ is the Minkowski metric (of ``mostly-plus'' signature) and $\epsilon$ is the antisymmetric Lorentz-invariant tensor such that $\epsilon^{012}=1$.  
The constant coefficients $g_{rs}$, 
$f_{rst}$  are defined such that they are symmetric under exchange of any two indices. We can view $g_{rs}$ as a metric on the `flavour'  space if we assume (as is the case  for the models of interest) that it is invertible. We need the $N=3$ case for TMG and the $N=4$ case for GMG.
The  time/space decomposition
\be
a^r = dt\,  a_0^r + d\xi^i a_i^r \quad (i=1,2)
\ee
yields a Lagrangian density of the form
\be\label{Lag3}
{\cal L} = -\frac{1}{2} \eta^{ab} \varepsilon^{ij} g_{rs}\,  a_{i\, a}^r \dot a_{j\, b}^s + a_{0\, a}^r \phi_r^a\, , 
\ee
where $\varepsilon^{ij} \equiv \varepsilon^{0ij}$. The phase space has dimension $6N$. The  time components $a_0^r$ impose $3N$ (primary)  
constraints $\phi_r=0$.

At this point we can anticipate how the Hamiltonian formulation will lead to the conclusion that GMG describes two local degrees of freedom, realized in the linear theory as two massive spin-2 modes. We can see from (\ref{secondary}) that  the equations $h_{[ij]}=0$ and $f_{[ij]}=0$ are additional secondary constraints, since they are  conditions on canonical variables that are not imposed by Lagrange multipliers but which are a consequence of the equations of motion.  Following Carlip's analysis of TMG \cite{Carlip:2008qh}, we are led  to consider
the modified Lagrangian 
\be\label{GMG+}
{\cal L}_{GMG^+} = {\cal L}_{GMG} + b_0\,  \varepsilon^{ij} h_{ij} + c_0 \, \varepsilon^{ij} f_{ij}
\ee
The phase space is spanned by the 2-space components of the four 3-vector one-form fields and hence has dimension per space point of $4\times 3\times 2=24$. The time components of the fields impose $4\times 3=12$ primary constraints, to which we must add the two secondary constraints imposed by the new variables $(b_0,c_0)$, making a total of $14$ constraints. Of these we expect six to be first class, corresponding to the built-in local Lorentz and 3-space diffeomorphism  invariance. The physical phase space will then have dimension, per space point, of  $24-14-6=4$,  corresponding (by our earlier definition) to two local degrees of freedom. 

We shall fill in the details later. First we wish to address two other important questions:
\begin{enumerate} 

\item Is the modified Lagrangian, with the extra secondary constraints, equivalent to the original unmodified Lagrangian? 

\item  Is the modified Lagrangian still invariant  under 3-space diffeomorphisms? Although these terms respect manifest 2-space diffeomorphism invariance they break manifest 3-space diffeomorphism invariance. 

\end{enumerate}
Notice that we could restore manifest 3-space diffeomorphism invariance by starting from the Lagrangian 3-form 
$L_{GMG} - b\, e\cdot h  - c\, e\cdot f$, so that  the secondary constraints are  imposed by the time components of the new 
one-form fields $b$ and $c$.  There is still the issue of equivalence of the new field equations to the original GMG field equations; in fact, equivalence is lost in making 
this modification but it can be restored (subject to a reservation to be discussed below)  by adding 
kinetic  terms for $(b,c)$ to arrive at the new Lagrangian 3-form 
\begin{eqnarray}\label{GMGbc}
L_{GMG^{++}} &=& -\sigma e_a R^a + \frac{1}{6} \Lambda_0 \epsilon^{abc} e_a e_b e_c  + h_a\left(T^a -be^a\right) + \frac{1}{2\mu} bdb + \frac{1}{m^2}c\left(db - e\cdot f\right)  \nonumber \\
&&+ \frac{1}{2\mu}\left[ \omega_a d\omega^a  + \frac{1}{3} \epsilon^{abc}\omega_a\omega_b\omega_c\right]-  \frac{1}{m^2} \left[f_a R^a + \frac{1}{2}\epsilon_{abc}e^af^bf^c \right]\, . 
\end{eqnarray}
Observe that 
\be
\left. {\cal L}_{GMG^{++}}\right|_{b_i=c_i=0}  = {\cal L}_{GMG^{+}} \, .
\ee
In other words, setting to zero the space components of the $(b,c)$ fields  yields the  minimal modification of (\ref{GMG+}). 

Whether we choose to consider the $GMG^+$ or the $GMG^{++}$ modification of  the GMG action, we still need to address
the issue of whether the modified field equations are equivalent to those of GMG. This is not obvious even for TMG; it was 
observed in \cite{Blagojevic:2008bn}  that Carlip's  Hamiltonian formulation of TMG (which amounts to using  $TMG^+$) was  incomplete because it did not include a proof  of equivalence to TMG. We provide a proof here that the field equations of both 
$TMG^+$ and  $TMG^{++}$ are equivalent to those of $TMG$.  As we shall see, the analogous equivalence proof  for GMG is more involved and does not  apply without qualification.  We shall begin our analysis by considering the 3-space covariant ``$++$'' modifications since this is simpler and 
the analysis is easily adapted to the non-covariant ``$+$'' modifications. 

\begin{itemize}

\item{TMG$^{++}$.}  By taking  $m\to\infty$ in (\ref{GMGbc}) we arrive at the following Lagrangian 3-form  in terms of the one-form fields of the HW model
\begin{eqnarray}\label{LagTMG}
L  &=&  -\sigma e_a R^a  + \frac{1}{6}\Lambda_0\,  \epsilon_{abc} e^a e^b e^c  +   h_a \left(T^a -be^a\right)  + \frac{1}{2\mu} bdb\nonumber \\
&&\ +\  \frac{1}{2\mu}\left[ \omega_a d\omega^a  + \frac{1}{3} \epsilon^{abc}\omega_a\omega_b\omega_c \right]\, . 
\end{eqnarray}
In the limit that $\sigma\to 0$ we must also set $\Lambda_0=0$ in order to get consistent field equations, and in this case we recover the HW model  after a rescaling of $h^a$. In other words, the model 
under consideration is essentially defined by an action that is the sum of the EC and HW actions. We shall now show that this
is another description of TMG. 

Consider first the $(h,\omega)$ equations, which are jointly equivalent to 
\be
T^a= be^a\, , \qquad \frac{1}{\mu} R^a + \epsilon^{abc} e_b h_c - \sigma b e^a = 0\, . 
\ee
These may be solved to give 
\be\label{solve1}
\omega_\mu{}^a = \omega_\mu{}^a(e,b)\, , \qquad h_{\mu\nu} = -\frac{1}{\mu}S_{\mu\nu}(e,b) + \sigma \epsilon_{\mu\nu\lambda} b^\lambda\, . 
\ee
The second of these equations implies that 
\be\label{antih}
h_{[\mu\nu]} +\frac{1}{\mu} \partial_{[\mu}b_{\nu]} =  \sigma\epsilon_{\mu\nu\lambda} b^\lambda\, . 
\ee
However, the $b$ equation is equivalent to the vanishing of the left hand side, from which we  deduce (assuming $\sigma\ne0$) that  $b=0$. Therefore,  the combined  $(h^a,\omega^a,b)$ field equations can be solved for these fields to  give
\be
\omega_\mu{}^a =\omega_\mu{}^a (e)\, , \qquad h_{\mu\nu} = -S_{\mu\nu}(e) \, , \qquad b_\mu=0\, . 
\ee
Back-substitution then yields the standard TMG action.  Alternatively, we may observe that since the field equations that follow from (\ref{LagTMG}) imply that $b=0$, the remaining equations are equivalent to those of TMG; specifically, those that follow from the 
TMG limit of the GMG Lagrangian 3-form (\ref{LagGMG2}). 

\item{GMG$^{++}$.}   We now turn to the general case of the Lagrangian 3-form (\ref{GMGbc}).  The $(h,f)$ equations
\be\label{hf}
T^a= be^a \, , \qquad R^a + \epsilon^{abc} e_b f_c = c e^a 
\ee
can be solved to give
\be
\omega_\mu{}^a = \omega_\mu{}^a(e,b) \, , \qquad f_{\mu\nu} = -S_{\mu\nu}(e,b) + \epsilon_{\mu\nu\rho} c^\rho\, , 
\ee
and hence 
\be\label{fb}
f_{[\mu\nu]}+ \partial_{[\mu}b_{\nu]} = - \epsilon_{\mu\nu\rho} c^\rho\, . 
\ee
On the other hand, the $c$-equation implies that $f_{[\mu\nu]}+ \partial_{[\mu}b_{\nu]}=0$, and hence that $c=0$. 

Let us now consider the $\omega$-equation; this is equivalent, given  (\ref{hf}) and $c=0$, to
\be\label{feq}
\frac{1}{m^2} Df^a -  \epsilon^{abc} e_b \left(h_c - \frac{1}{\mu} f_c\right) + \sigma b e^a =0\, ,  
\ee
which has the following solution\footnote{It is useful to use here the fact that $(D+b)f^a$ is $b$-independent given (\ref{hf}). To see this, compare with the equation of motion (\ref{HWeq}) of the HW model.}:
\begin{eqnarray}\label{hwithb}
m^2\left(h_{\mu\nu} - \frac{1}{\mu} f_{\mu\nu}\right) &=&   C_{\mu\nu}(e) + \frac{1}{2} g_{\mu\nu} \left(\epsilon^{\rho\sigma\tau}b_\rho\partial_\sigma b_\tau\right)
+ \epsilon_\nu{}^{\tau\lambda} b_\tau S_{\lambda\mu}(e)\nonumber \\
&&\qquad +\,  \epsilon_\nu{}^{\tau\lambda} b_\tau \bar D_\lambda b_\mu + \frac{1}{2} \epsilon_{\mu\nu\lambda} b^\lambda \left(b^2-2\sigma m^2\right)\, . 
\end{eqnarray}
However,  the $b$ and $c$ equations combined tells us, given  $c=0$, that the antisymmetric part of the left hand side is zero.  
So the antisymmetric part of the right hand side is zero, and this is equivalent to the equation
\be\label{GMGbeq}
\Xi^{\mu\nu}b_\nu=0\, , \qquad \Xi^{\mu\nu} \equiv G^{\mu\nu} - \bar D^\mu b^\nu - g^{\mu\nu}\left(2\sigma m^2 + \bar D\cdot b- b^2 \right)\, . 
\ee
Although $b=0$ solves this equation,  there are other possible solutions,   so  GMG$^{++}$ is  not strictly equivalent  to GMG. Instead, it appears that GMG is equivalent to one ``branch'' of the model defined by the GMG$^{++}$ action. 

This branch equivalence goes beyond the statement that the field equations of GMG$^{++}$ reduce to those of GMG when we choose the $b=0$ solution of 
(\ref{GMGbeq}).  An analysis of fluctuations about any solution of  GMG$^{++}$ with $b=0$ leads to the linear equation 
$\left[G_{\mu\nu} - 2\sigma m^2 g_{\mu\nu}\right]\delta b^\nu=0$, which  implies that $\delta b=0$, generically, and hence that a linear stabilty analysis for solutions of GMG$^{++}$ with $b=0$ is equivalent to a linear stability analysis in the context of  GMG.  An exception to this state of affairs occurs when we consider  fluctuations about a maximally-symmetric vacuum  with  $G_{\mu\nu} = 2\sigma m^2 g_{\mu\nu}$. This  is precisely the case in which the massive gravitons become  ``partially massless'' \cite{Bergshoeff:2009hq}. In this case $b$ is undetermined at the linear level, so there must be an ``accidental'' gauge invariance of the linear theory that allows it to be ``gauged away''. At the non-linear level it is still present but we may also still choose the $b=0$ solution.

\end{itemize}

Recall that we introduced the  GMG$^{++}$ Lagrangian 3-form (\ref{GMGbc}) in an attempt to restore the manifest 3-covariance that is lost 
when considering the minimal modification of GMG$^+$. However, if we were to take this as our starting point for a Hamiltonian formulation we would need to include the non-covariant conditions $b_i=c_i=0$ as new secondary constraints because these are constraints on canonical variables that are either implied by the field equations or imposed consistently with them. As a check we observe that we would then have a phase space of dimension $4\times 3\times 2 + 2\times 2=28$, per space point, and  $12+2+4=18$ constraints. Given that  six of these constraints are first class we then get a physical phase space of  dimension $28-18-6=4$, as expected. However, this would give us a non-minimal Hamiltonian formulation that has no obvious advantage over the minimal formulation provided by GMG$^+$.  

We therefore return to the  GMG$^+$ Lagrangian density of  (\ref{GMG+}). Although the inclusion of the secondary constraint terms, with Lagrange multipliers $b_0$ and $c_0$,  breaks manifest  3-space covariance, the field equations will still be 3-space covariant if they can be shown to be equivalent to those of GMG, so we now need to address this issue, which we  can do by adapting our analysis above  for GMG$^{++}$. 
We shall again consider the TMG case separately.

\begin{itemize} 

\item{TMG$^+$.} As $b_i=0$ the $b$ equation is now $h_{[ij]}=0$. The $(h^a,\omega^a)$ equations are just those of  TMG$^{++}$  but with $b_i=c_i=0$. This applies, in particular, to  (\ref{antih}) from which we see that $\sigma b_0$  is proportional to $h_{[ij]}$, and is therefore zero. As long as $\sigma$ is non-zero this implies that $b_0=0$, and hence $b=0$. Using this, the  remaining equations become those of TMG. 

\item{GMG$^+$.} By setting $b_i=c_i=0$ in (\ref{fb}) we deduce that $f_{[ij]}$ is proportional (for finite $m$) to $c_0$, but the $c_0$ equation is $f_{[ij]}=0$, so we deduce that $c_0=0$, and hence that $c=0$. So now we have equations that are equivalent to those of GMG except for possible $b_0$ dependence. 

The combined $(b_0,c_0)$ equations imply the symmetry of $\mu h_{ij} -f_{ij}$.   Using this in (\ref{hwithb}) and setting $b_i=0$ we arrive at the equation 
\be\label{b_0eq}
\left. \Xi^{00}\right|_{b_i=0} b_0 =0\, , 
\ee
where $\Xi$ is the tensor of (\ref{GMGbeq}). This equation has $b_0=0$ as a solution, and choosing this solution we get equivalence with GMG. 

\end{itemize}
We see that the branching of possibilities for $b$ that we found in our previous analysis of GMG$^{++}$ has not been  entirely eliminated. 
The equation (\ref{b_0eq}) has $b_0=0$ as one solution but there are other possibilities, at least one of which coincides with the $b_0=0$ solution at ``partially massless'' vacua.  A similar branching of possibilities at these vacua was found in the analysis of \cite{Afshar:2011qw}. From our perspective, it appears necessary to insist on $b_0=0$ because otherwise the field equations are not those of GMG.

\section{Hamiltonian formulation} 
\setcounter{equation}{0}

Following the discussion of the previous section, we  take as our GMG Lagrangian 3-form 
\begin{eqnarray}\label{LagGMG+}
L_{GMG^+} &=& -\sigma e_a R^a  + \frac{1}{6} \Lambda_0 \epsilon^{abc} e_a e_b e_c 
+ h_a T^a  + \frac{1}{2\mu}\left[ \omega_a d\omega^a  + \frac{1}{3} \epsilon^{abc}\omega_a\omega_b\omega_c\right]\nonumber \\
&&\  - \ \frac{1}{m^2} \left[ f_a R^a + \frac{1}{2}\epsilon^{abc} e_a f_b f_c \right] - \bar b \, e\cdot h - \frac{1}{m^2}\bar c\,  e\cdot f
\, , 
\end{eqnarray}
where
\be
\bar b = b_0 dt\, , \qquad \bar c= c_0 dt\, . 
\ee
The $(b_0,c_0)$ fields are now Lagrange multipliers for the constraints
\be
h_{[ij]}=0\, , \qquad f_{[ij]}=0\, . 
\ee
We shall call these the ``secondary'' constraints since they are secondary in the context of the GMG action (\ref{LagGMG2}).

The Lagrangian density still  takes the general form (\ref{TopLag}) except that we must now add the secondary constraints. 
After  a time/space decomposition we then arrive the Lagrangian density
\be\label{Lag3+}
{\cal L}_+ = -\frac{1}{2}  \varepsilon^{ij} g_{rs} a_i^r \cdot \dot a_j^s + a_0^r \cdot \phi_r + b^I_0 \psi_I\, , 
\ee
with $I=1,\dots n$, and  $n=1$ for TMG and $n=2$ for GMG. The constraint functions are 
\be\label{confun}
\phi_r = \varepsilon^{ij} \left( \partial_i a_j^s g_{rs} + \frac{1}{2}  f_{rst} a_i^s \times  a_j^t \right)\, , \qquad \psi_I = \frac{1}{2} f_{I,pq}\Delta^{pq} \, , 
\ee
where $f_{I, pq} = -f_{I,qp}$ is a new set of constant coefficients,  and 
\be
\Delta^{pq} = \varepsilon^{ij} a_i^p \cdot a_j^q\, . 
\ee
Observe that $\Delta^{pq}=-\Delta^{qp}$. 
The quadratic term of (\ref{Lag3+}) gives us the Poisson brackets
\be\label{PBs}
\left\{ a_{i\, a}^r (\xi), a_{j\, b}^s (\zeta)\right\}_{PB} = \eta_{ab} \, g^{rs} \varepsilon_{ij} \, \delta^{(2)}\left(\xi-\zeta\right)\, , 
\ee
which we may use to compute the matrix of Poisson brackets of the constraint functions. To this end, it is 
convenient to first define
\be\label{psipsi}
\phi(\alpha)=  \int \!d^2\xi \, \alpha_a^r(\xi) \phi_r^a(\xi)\, ,
\ee
where the test functions $\alpha_a^r$ are arbitrary except that we choose them such that no surface terms arise upon integration by parts. A calculation using (\ref{PBs}) yields the result
\begin{eqnarray}
\left\{\phi(\alpha),\phi(\beta)\right\}_{PB} &=&  \phi([\alpha,\beta]) + f^t{}_{q[r}f_{s]pt}\int d^2\xi\,  \alpha^r\cdot \beta^s \Delta^{pq} \nonumber \\
&& + \,  2 f^t{}_{r[s} f_{q]pt}\int\! d^2\xi \, \alpha^r_a \beta^s_b  \left(V^{ab}\right)^{pq}  \, , 
\end{eqnarray}
where
\be
\left[ \alpha,\beta\right]_t^c = \epsilon^{abc} \alpha_a^r \beta_b^s f_{rst}\, , \qquad 
V_{ab}^{pq} = \varepsilon^{ij} a_{i\, a}^{p} a_{j\, b}^{q} \, . 
\ee
As we discuss in the following subsections, it turns out that  for the models of interest   the  $\Delta$ terms in these Poisson brackets  are all zero as 
a consequence of the secondary constraints. It follows that, on the constraint surface, 
\be\label{Pmatrix}
\left\{\phi_r^a(\xi),\phi_s^b(\zeta)\right\} = P_{rs}^{ab}(\xi-\zeta)\, , 
\ee
where
\be
P_{rs}^{ab}(\xi-\zeta) =  f^t{}_{r[s} f_{q]pt} \left(V^{ab}\right)^{pq} \,  \delta^{(2)}(\xi-\zeta)\,  . 
\ee
This $3N\times 3N$ matrix  $P$  plays a crucial role in the analysis to follow.  However, we will also need to take into account the Poisson brackets  of the primary with the secondary constraints. A calculation for the generic model shows that 
\be\label{phipsi}
\left\{\phi(\alpha), \psi_I\right\}_{PB} = \varepsilon^{ij} \left[\partial_i \alpha^r a_j^q f_{I,rq} - \alpha^r a_i^s\times a_j^q f_{rs}{}^p f_{I,pq}\right]\, . 
\ee
For NMG and GMG we also need the Poisson bracket of the two secondary constraints; we shall see that these two constraints are in involution. 

\subsection{TMG} 

Separating the quadratic from the cubic terms in (\ref{LagTMGs}), we find that 
\begin{eqnarray}\label{LagTMG2}
L_{TMG} &=&  -\sigma e_a d\omega^a  + h_a de^a  + \frac{1}{2\mu}\omega_a d\omega^a  \nonumber \\
&& +\ \epsilon_{abc}\left\{ h^a \omega^b e^c -  \frac{\sigma}{2} e^a \omega^b\omega^c + \frac{1}{6\mu}\omega^a\omega^b\omega^c + \frac{\Lambda_0}{6} e^a e^b e^c \right\} \, . 
\end{eqnarray}
We can simplify the quadratic term by setting
\be
\omega^a= \Omega^a + \sigma\mu\,  e^a\, , \qquad h^a = k^a + \frac{1}{2}\sigma^2\mu\,  e^a\, . 
\ee
In terms of  $(e,k,\Omega)$ the Lagrangian 3-form is
\be\label{newTMG}
L_{TMG} =  k_a de^a + \frac{1}{2\mu} \Omega_a d\Omega^a  +\epsilon_{abc} \left[ k^a\Omega^b e^c + \frac{1}{6\mu}\Omega^a\Omega^b\Omega^c + \sigma\mu\,  k^a e^b e^c + \frac{1}{6}\tilde\Lambda_0 \, e^a e^b e^c \right]\, , 
\ee
where
\be
\tilde\Lambda_0 = \Lambda_0 + \sigma^3\mu^2\, . 
\ee
Apart from differences in notation, this result differs from the analogous result of the HW model only in the $\sigma$- and 
$\Lambda_0$-dependent cubic terms.  

We are now in a position to write down a Hamiltonian form of the action, by performing a time/space split in (\ref{newTMG}) and adding the secondary constraint. This gives us, 
\be
{\cal L}_{TMG^+} = - \varepsilon^{ij} \left\{ k_i \cdot \dot e_j + \frac{1}{2\mu} \omega_i \cdot\dot\omega_j\right\} + e_0\cdot \phi_e
+ \omega_0\cdot \phi_\omega + k_0 \cdot \phi_k +  b_0 \Delta^{ek}\, , 
\ee
where
\begin{eqnarray}
\phi_\omega &=&  \varepsilon^{ij} \left\{\frac{1}{\mu}\left[\partial_i \omega_j + \frac{1}{2}\omega_i\times \omega_j\right] 
+ e_i\times k_j \right\} \nonumber \\
\phi_e &=& \varepsilon^{ij} \left\{ D_i k_j + 2\sigma\mu\,  k_i \times e_j + \frac{1}{2} \tilde\Lambda_0 \, e_i \times e_j\right\} \nonumber \\
\phi_k &=& \varepsilon^{ij} \left\{ D_i e_j + \sigma\mu\, e_i \times e_j \right\}\, . 
\end{eqnarray} 
These constraint functions are, of course, just the specialization to the case in hand of those given by the general formula (\ref{confun}).  We see from this result that the phase space, spanned by the space components of the Lorentz 
3-vectors $(\omega,e,k)$ has dimension $18$ per space point, and that there are $10$ constraints. 

We actually have no need for the above explicit expressions for the primary constraint functions $\phi_r$ since we may use the general result  for their Poisson brackets  that we have already computed in terms of the various coefficients that define the model. To do this we first read off from (\ref{newTMG}) the non-zero components of $g$ and $f$, which are
\be
g_{ek} = 1\, , \qquad g_{\Omega\Omega}= \mu^{-1}\, , 
\ee
and 
\be
 f_{k\Omega e} = 1\, , \qquad f_{\Omega\Omega\Omega} = \mu^{-1}\, , \qquad f_{kee} = 2\sigma\mu \, , \qquad 
f_{eee}= \tilde\Lambda_0\, , 
\ee
from which it follows that
\begin{eqnarray}
f^{\Omega}{}_{ke} &=&\mu\, , \qquad  f^\Omega{}_{\Omega\Omega}= f^e{}_{\Omega e} = f^k{}_{\Omega k} =1\, , \nonumber\\
f^k{}_{ee} &=& \tilde\Lambda_0\, , \qquad  f^e{}_{ee} = f^k{}_{ke} = 2\sigma\mu\, . 
\end{eqnarray}

Using these results we find that the $9\times 9$ $P$-matrix of (\ref{Pmatrix}) takes the following  form in the  $(\Omega,k,e)$ basis:
\be\label{LLinv}
\left(P_{ab}\right)_{rs}= \left(\begin{array}{cc} 0 & 0 \\ 0&  Q\end{array}\right)
\ee
where $Q$ is the antisymmetric $6\times 6$ matrix \footnote{It is  antisymmetric because $-\left(V_{ab}^{ek}\right)^T = V_{ab}^{ke}$.}
\be
Q= \mu^{-1}\delta^{(2)}(\xi-\zeta) \left(\begin{array}{cc}  -V_{ab}^{ee} & V_{ab}^{ek} \\  V_{ab}^{ke} & -V_{ab}^{kk} \end{array}\right)\, . 
\ee
Note that this matrix  is independent of both $\sigma$ and $\Lambda_0$.  
The zeros of the first row and column of $P$ (actually 3 rows and 3 columns because we suppress Lorentz indices) are 
expected from the built-in local Lorentz invariance, which ensures that the Poisson bracket  of $\phi_\Omega$ with any other constraint is zero on the constraint surface, i.e. that  the constraints $\phi_\Omega^a$ are first-class.  
We may use this built-in local Lorentz invariance to choose a local frame for which $e^a_1 = (0\,1\,0),\,e^a_2 = (0\,0\,1)$, in which case the secondary  constraint  implies  that $k^2_1 = k^1_2$. It can then be easily verified using Mathematica that the matrix $Q$ (and hence $P$) has rank $2$. 

However, we still have to take into account the one secondary constraint, with constraint function $\psi= \Delta^{ek}$. 
From the general result (\ref{phipsi}) we find, in this instance,  that 
\be
\left\{\phi(\alpha),\psi\right\}_{PB} = \varepsilon^{ij} \left(D_i\alpha^e k_j - D_i\alpha^k e_j\right) 
+ \left(2\sigma\mu \, \alpha^k  +  \tilde\Lambda_0 \alpha^e\right) \varepsilon^{ij} e_i \times e_j 
\ee
where we use the shorthand
\be
D_i \alpha^r a_j^s \equiv \partial_i \alpha^r a_j^s - \alpha^r \Omega_i \times a_j^s\, . 
\ee
The absence of any term involving $\alpha^\Omega$ is expected from the fact that the secondary constraint function is a Lorentz scalar,  but  both $\phi_e$ and $\phi_k$  have non-zero Poisson brackets  with $\psi$:
\begin{eqnarray}\label{vs}
\!\!\!\!\!\!\!\!\left\{\phi_e(\xi),\psi(\zeta)\right\}_{PB} &=& -\varepsilon^{ij}\left\{k_j\,  \partial_i \delta^{(2)}(\xi-\zeta)  - \left[\Omega_i\times k_j - \tilde\Lambda_0 \, e_i\times e_j\right]\delta^{(2)}(\xi-\zeta)\right\} \nonumber \\
\!\!\!\!\!\!\!\!\left\{\phi_k(\xi),\psi(\zeta)\right\}_{PB} &=&  \varepsilon^{ij}\left\{ e_j\,  \partial_i \delta^{(2)}(\xi-\zeta) + \left[ \Omega_i\times e_j + 2\sigma\mu \, e_i\times e_j \right] \delta^{(2)}(\xi-\zeta)\right\} \, . 
\end{eqnarray}
This shows that the  $10\times 10$ matrix $\bP$ of Poisson brackets of constraints takes the form
\be\label{bPform}
\bP = \left(\begin{array}{cc}  0 & 0 \\ 0 & \bQ  \end{array}   \right) \, , 
\ee
where $\bQ$ is a  $7\times 7$ antisymmetric matrix of the form 
\be
\bQ = \left(\begin{array}{cc}  Q & v  \\   -v^T &0 \end{array}\right) \, , 
\qquad v= \left(\begin{array}{c}\left\{\phi_k, \psi\right\}_{PB} \\  \left\{\phi_e, \psi\right\}_{PB}\end{array}\right)\, .
\ee
All dependence on $\sigma$ and $\Lambda_0$ enters through  the column vector $v$. If  this column vector is  in the column space of $Q$ then the rank of $\bQ$ equals the rank of $Q$, i.e $2$. This would imply that there are $8$ first-class constraints and hence $18-10-8=0$ local degrees of freedom. 
This special case is realized if and only if $\sigma=\Lambda_0=0$, as expected because this is the limit in which TMG degenerates to conformal gravity, which has no local degrees of freedom. The $8$ gauge invariances are what remains of the 
local conformal invariance in the gauge in which $b_i=0$.  In all other cases $v$ is not in the column space of $Q$, and  the rank of $\bQ$ is then $2$ greater than the rank of $Q$, i.e. $4$. This implies that there are six  first-class constraints, per space point, corresponding to  six gauge invariances that can be identified  as those of  3-space diffeomorphisms and local Lorentz invariance \cite{Carlip:2008qh}. The dimension, per space point, of the physical phase space is therefore $18-10-6=2$, as expected for TMG.

\subsection{NMG} 

For a reason that will become clear, we develop the Hamiltonian formalism for NMG separately from that of GMG. Taking the 
$\mu\to\infty$ limit in (\ref{LagGMG2}) we arrive at  the  NMG Lagrangian 3-form: 
\begin{eqnarray}\label{LagNMG}
L_{NMG} &=& -\sigma e_a R^a  + \frac{1}{6} \Lambda_0 \epsilon^{abc} e_a e_b e_c 
+ h_a T^a  - \frac{1}{m^2} \left[ f_a R^a + \frac{1}{2}\epsilon^{abc} e_a f_b f_c \right] \, . 
\end{eqnarray}
Separating the quadratic and cubic terms, and then simplifying the former by defining the new variable 
\be\label{ftopi}
\pi^a = -\frac{1}{m^2} f^a - \sigma e^a \, ,  
\ee
we arrive at the Lagrangian 3-form 
\begin{eqnarray}\label{NMGnewbasis}
L_{NMG} &=&  h_a de^a  + \pi_a d\omega^a + 
\epsilon_{abc} \left[ h^a\omega^b e^c + \frac{1}{2} \pi^a \omega^b\omega^c \right] \nonumber \\
&&+\  \epsilon_{abc} \left[ - \frac{m^2}{2} e^a \pi^b \pi^c - m^2\sigma e^a e^b \pi^c + \frac{1}{6} \hat \Lambda_0 - e^a e^b e^c \right] \, ,   
\end{eqnarray}
where
\be
\hat\Lambda_0 = \Lambda_0-3m^2\sigma\, . 
\ee
Making a time/space split, and then adding the two secondary constraints, we arrive at the  NMG Lagrangian density in Hamiltonian form
\begin{eqnarray}
{\cal L}_{NMG^+} &=& - \varepsilon^{ij} \left\{ h_i \cdot \dot e_j + \pi_i \cdot \dot \omega_j\right\} + b_0\Delta^{eh} + c_0 \Delta^{e\pi} \nonumber \\
&&+\  \omega_0 \cdot \phi_\omega + e_0
\cdot \phi_e + h_0 \cdot \phi_h + \pi_0 \cdot \phi_\pi \, , 
\end{eqnarray}
where
\begin{eqnarray}
\phi_\omega &=& \varepsilon^{ij} \left\{ D_i \pi_j + e_i \times h_j\right\} \, , \qquad
\phi_h =  \varepsilon^{ij} D_i e_j \, , \nonumber \\
\phi_e &=&  \varepsilon^{ij} \left\{ D_i h_j - \frac{m^2}{2} \pi_i\times\pi_j -2m^2\sigma\,  e_i \times \pi_j 
+ \frac{1}{2}\hat\Lambda\,  e_i\times e_j \right\}\nonumber \\
\phi_\pi &=& \varepsilon^{ij} \left\{ \left[D_i \omega_j + \frac{1}{2}\omega_i\times\omega_j\right] - m^2 e_i\times \pi_j - m^2\sigma e_i\times e_j \right\}\, .  
\end{eqnarray}
The phase space now has dimension, per space point of $24$ but there are a total of $14$ constraints. Our next task is to determine 
the dimension of the subspace of first-class constraints and hence the number of gauge invariances. 

As for TMG, we do not need to use directly the above explicit  expressions for the primary constraint functions  because we may instead use the general result for the Poisson brackets of the constraint functions that we computed earlier. For this we need the expressions for the coefficients $g$ and $f$ in the  $(\omega,\pi,h,e)$ basis, which we may read off from (\ref{NMGnewbasis}). The non-zero components of $g$ and $f$ are
\begin{eqnarray}
f_{eh}&=&  g_{\pi\omega}=1 \, , \qquad 
f_{h\omega e} = f_{\pi\omega\omega}=1 \nonumber \\
f_{e\pi\pi} &=&-m^2 \, , \qquad f_{ee\pi}= -2m^2\sigma \, , \qquad f_{eee}= \hat\Lambda_0\, , 
\end{eqnarray}
from which it follows that 
\begin{eqnarray}
f^e{}_{\omega e}&=& f^h{}_{\omega h} = f^\omega{}_{\omega\omega} = f^\pi{}_{eh} =f^\pi{}_{\pi\omega}=1\, , \qquad f^h{}_{\pi\pi} = f^\omega{}_{e\pi} = -m^2 \, , \nonumber \\
f^h{}_{e\pi}&=& f^\omega{}_{ee} = -2m^2\sigma\, , \qquad f^h{}_{ee} = \hat\Lambda_0 \, . 
\end{eqnarray}
Using these results we find that the $P$ matrix of (\ref{Pmatrix}) in the  $(\omega,\pi,h,e)$ basis again takes the form
(\ref{LLinv}) but now with a  $9\times 9$ antisymmetric submatrix  $Q$;  suppressing Lorentz indices, we have 
\be
Q=  m^2\delta^{(2)}(\xi-\zeta) \left(\begin{array}{ccc}  0 & V^{ee} & -V^{eh}  \\ 
V^{ee}& 0 & -V^{e\pi} \\  -V^{he} & -V^{\pi e}& V^{h\pi} + V^{\pi h} \end{array}\right)\, . 
\ee
Note that this matrix is independent of both $\sigma$ and $\Lambda_0$.  It is antisymmetric because, for example, the transpose of $V^{ee}$ is $-V^{ee}$. For the same choice of frame for  $e_i{}^a$ that we used  for TMG, we have  $h_1^2 = h_2^1$ and $\pi_1^2 = \pi_2^1$, and a Mathematica calculation can then be used to show that $Q$ has rank $4$.

Now we must take into account the two secondary constraints, with constraint functions
\be
\psi_1 = \Delta^{eh}\, , \qquad \psi_2= \Delta^{e\pi}
\ee
It is easily verified that $\left\{\psi_1,\psi_2\right\}_{PB} \propto \psi_2$, so this PB is zero on the constraint surface. 
The non-trivial Poisson brackets  are
\begin{eqnarray}\label{vsNMG1}
\left\{\phi(\alpha),\psi_1 \right\}_{PB} &= &  \varepsilon^{ij} \left[D_i \alpha^e h_j - D_i\alpha^h e_j 
- m^2\alpha^\pi\pi_i \times e_j \right] \nonumber \\
&&+\ \varepsilon^{ij} \left[ \left(\alpha^e+\alpha^\pi\right) \hat \Lambda \, e_i\times e_j - 2m^2\sigma\, \alpha^e \, \pi_i\times e_j\right]\, , 
 \nonumber \\
\left\{\phi(\alpha),\psi_2 \right\}_{PB} &= & \varepsilon^{ij} \left[D_i \alpha^e\pi_j - D_i\alpha^\pi e_j + \alpha^e \, h_i\times e_j +\alpha^h \, e_i\times e_j\right]\,  
\end{eqnarray}
where we again use a shorthand notation:
\be
D_i \alpha^r a_j^s \equiv \partial_i \alpha^r a_j^s - \alpha^r \omega_i \times a_j^s\, . 
\ee

This gives us a $14\times 14$ $\bP$-matrix of the general form (\ref{bPform}) but now with an $11\times 11$ antisymmetric sub-matrix $\bQ$ of the form
\be\label{bQformNMG}
\bQ = \left(\begin{array}{ccc}  Q & v_1 & v_2 \\ -v_1^T & 0 & 0 \\ -v_2^T & 0 & 0 \end{array}
 \right) \, , \qquad
v_I = \left(\begin{array}{c} \left\{\phi_\pi,\psi_I\right\}_{PB} \\  \left\{\phi_h,\psi_I\right\}_{PB} \\  
\left\{\phi_e,\psi_I\right\}_{PB} \end{array}\right)\, ,  
\ee
where the Poisson brackets $\left\{\phi_r,\psi_I\right\}_{PB}$ can be read off from (\ref{vsNMG1}).  Observe that the components of these column vectors are sums of terms that are either linear or quadratic in canonical variables, and that all dependence on $\sigma$ and $\Lambda_0$ is contained in the quadratic terms. 

For a matrix $\bQ$ of the above form, its rank  equals the rank of $Q$  (i.e. $4$) if both $v_1$ and $v_2$ are in the column space of $Q$, and it equals the rank of $Q$ plus $4$ (i.e. $8$) if  $v_1$ and $v_2$ are linearly independent and no linear combination of them is in the column space of $Q$. In all other cases the rank of $\bQ$ is the rank of $Q$ plus $2$ (i.e. $6$).  If the quadratic terms of $v_I$ were absent, then both these vectors would be in the column space of $Q$, so that $\bQ$ would have rank $4$. However, the most we can do to eliminate these quadratic terms is to set  $\sigma=\Lambda_0=0$ and  this still leaves some quadratic terms, which  are sufficient to ensure that the column vectors $v_I$ are both  independent and that no linear combination of them is  in the column space of $Q$,  so the rank of  $\bQ$ is $8$  {\it independently of the values of $\sigma$ or $\Lambda_0$}.  This means that $6$ of the $14$ constraints are first-class, as expected, and hence that the dimension of the physical phase space, per space point, is  $24-14-6=4$. This is consistent with the linearized analysis of the generic NMG model, which shows that there are two propagating modes, and with the Hamiltonian results of  \cite{Blagojevic:2010ir},  but it also applies in the $\sigma=\Lambda_0=0$ limit that yields ``massless NMG'', and in that case it is not consistent with the linearized analysis of  \cite{Deser:2009hb}.  We conclude  that ``massless NMG'' suffers from a linearization instability. 

\subsection{GMG} 

For GMG we proceed initially as for NMG, separating the quadratic from the cubic terms of the GMG Lagrangian 3-form 
 (\ref{LagGMG2}) and then  making the change of variable (\ref{ftopi}) to get to 
\begin{eqnarray} 
L_{GMG} &=&  h_a de^a + \frac{1}{2\mu} \omega_a d\omega^a + \pi_a d\omega^a + 
\epsilon_{abc} \left[ h^a\omega^b e^c + \frac{1}{6\mu} \omega^a\omega^b\omega^c + \frac{1}{2} \pi^a \omega^b\omega^c \right] \nonumber \\
&&+\  \epsilon_{abc} \left[ - \frac{m^2}{2} e^a \pi^b \pi^c - m^2\sigma e^a e^b \pi^c + \frac{1}{6} \hat\Lambda_0\,   e^a e^b e^c \right] \, . 
\end{eqnarray}
To further simplify the quadratic term we now set 
\be
\omega^a = \Omega^a - \frac{\mu}{m^2} \pi^a \, ,  
\ee
where $\Omega^a$ is a new independent connection. This gives us the Lagrangian 3-form 
\begin{eqnarray}\label{GMGnewbasis}
L_{GMG} &=& h_a de^a + \frac{1}{2\mu} \Omega_a d\Omega^a - \frac{\mu}{2} \pi_a d\pi^a
+ \epsilon_{abc}\left[h^a \Omega^b e^a+  \frac{1}{6\mu} \Omega^a\Omega^b\Omega^c - \frac{\mu}{2} \pi^a \Omega^b\pi^c \right] \nonumber \\
&& \!\!\!\!  \!\!\!\!  \!\!\!\!  \!\!\!\!  \!\!\!\!  \!\!\!\!  \!\!\!\!  \!\!\!\! 
+ \ \epsilon_{abc} \left[ \frac{\mu^2}{3}\pi^a\pi^b\pi^c  - \mu h^a \pi^b e^c 
- \frac{m^2}{2} e^a\pi^b\pi^c -m^2\sigma e^a e^b \pi^c + \frac{1}{6}\hat \Lambda_0 \,  e^a e^b e^c \right] \, . 
\end{eqnarray}
Observe that all $\Omega$-terms in the cubic term covariantize the quadratic terms with respect  to local Lorentz transformations, so local Lorentz invariance is still manifest.   The `flavour' space metric  is simple in the new $(\Omega, \pi, e,h)$ basis, but it is no longer simple to consider the $\mu\to\infty$ limit that yields NMG. This is why we first dealt separately with the NMG case; having done so we may now assume that $\mu$ is finite.  

Making a time/space split and adding the two secondary constraints, we  arrive at the following Hamiltonian form of the GMG Lagrangian density:
\begin{eqnarray}
{\cal L}_{GMG+} &=& - \varepsilon^{ij}\left\{ h_i \cdot \dot e_j + \frac{1}{2\mu} \Omega_i \cdot \dot\Omega_j - \frac{\mu}{2} \pi_i \cdot \dot\pi_j \right\} - b_0 \Delta^{eh} -c_o \Delta^{e\pi} \nonumber\\
&&+\  \omega_0 \cdot \phi_\omega + e_0
\cdot \phi_e + h_0 \cdot \phi_h + \pi_0 \cdot \phi_\pi \, , 
\end{eqnarray}
where
\begin{eqnarray}
\phi_\Omega &=& \varepsilon^{ij} \left\{ \frac{1}{2\mu} \left[\partial_i\Omega_j + \frac{1}{2} \Omega_i\times\Omega_j \right] + e_i\times h_j - \frac{\mu}{2} \pi_i\times\pi_j \right\} \, , \nonumber \\
\phi_h &=& \varepsilon^{ij} \left\{ D_i e_j - \mu \pi_i\times e_j\right\}\, , \nonumber \\
\phi_e &=&  \varepsilon^{ij} \left\{ D_i h_j - \mu h_i\times \pi_j -  \frac{m^2}{2} \pi_i\times\pi_j -2m^2\sigma\,  e_i \times \pi_j 
+ \frac{1}{2}\hat\Lambda\,  e_i\times e_j \right\}\nonumber \\
\phi_\pi &=& \varepsilon^{ij} \left\{ -\frac{\mu}{2} D_i \pi_j + \mu^2 \pi_i\times\pi_j - m^2 e_i\times \pi_j - m^2\sigma e_i\times e_j \right\}\, .  
\end{eqnarray}

Again, we do not need to use these expressions directly because we may instead use the general result for the Poisson brackets of the constraint functions that we computed earlier in terms of the coefficients $g$ and $f$ that define the model. From (\ref{GMGnewbasis}) we see that the non-zero coefficients in  the  $(\Omega,e,\pi,h)$ basis are
\be
g_{eh} =1 \, , \qquad g_{\Omega\Omega} = \mu^{-1} \, , \qquad g_{\pi\pi} =- \mu\, , 
\ee
and that the non-zero components of $f_{rst}$ are
\begin{eqnarray}
f_{h\Omega e} &=&1\, , \qquad f_{\Omega\Omega\Omega} = \mu^{-1}\, , 
\qquad f_{\pi\pi\Omega} = - \mu \, , \nonumber \\
f_{\pi\pi\pi} &=& 2\mu^2\, , \qquad f_{h\pi e} = -\mu \, , \nonumber \\
f_{e\pi\pi} &=& -m^2\, , \qquad  f_{e e\pi} = -2m^2\sigma\, , \qquad f_{e e e} = \hat\Lambda_0\, . 
\end{eqnarray}
It follows that
\begin{eqnarray}
f^e{}_{\Omega e} &=&  f^h{}_{\Omega h} = f^\pi{}_{\Omega\pi} = f^\Omega{}_{\Omega\Omega}=1 
\, , \qquad  f^\Omega{}_{he} = \mu \, , \qquad f^\Omega_{\pi\pi} = -\mu^2\, , \nonumber\\
f^\pi{}_{h e}&=&1\, , \qquad 
f^\pi{}_{\pi\pi} = -2\mu\, , \qquad f^\pi{}_{e \pi} = m^2/\mu\, , \qquad 
f^\pi{}_{e e} = 2m^2\sigma/\mu\, , \nonumber \\
f^e{}_{\pi e} &=& f^h{}_{\pi h} =-\mu\, , \qquad 
f^h{}_{\pi\pi} = -m^2\, , \qquad f^h{}_{e\pi} = -2m^2\sigma\, , \nonumber \\
 f^h{}_{e e}&=& \hat\Lambda_0 \, . 
\end{eqnarray}
Using these results we find that the $P$ matrix of (\ref{Pmatrix}) in the  $(\Omega,\pi,h,e)$ basis again takes the form
(\ref{LLinv}) but the $9\times 9$ antisymmetric submatrix  $Q$ is now
\be
Q= \frac{m^2}{\mu} \delta^{(2)}(\xi-\zeta)\left(\begin{array}{ccc}   m^2V^{ee} & \mu V^{ee} & -\mu V^{eh} - m^2 V^{e\pi} \\  
\mu V^{ee} & 0 & -\mu V^{e\pi} \\  
-\mu V^{he} - m^2 V^{\pi e} & -\mu V^{\pi e} & \mu(V^{\pi h} + V^{h \pi}) + m^2V^{\pi\pi} \\
\end{array}\right)\, . 
\ee
Once again, this matrix is independent of both $\sigma$ and $\Lambda_0$. 
Using the same choices for $e_i^a$ as before, we can now use Mathematica to evaluate the rank of this matrix. We assume that neither  $m^2$ nor $\mu$ is  zero or infinity. The result is that $Q$ has rank $4$. 

Now we consider the secondary constraints; the constraint functions are  again
\be
\psi_1 = \Delta^{eh}\, , \qquad \psi_2= \Delta^{e\pi}\, , 
\ee
and it is again  straightforward to verify that $\left\{\psi_1,\psi_2\right\}_{PB} \propto \psi_2$, so this PB is 
is zero on the constraint surface. Next, we compute
\begin{eqnarray}\label{vsGMG}
\left\{\phi(\alpha),\psi_1 \right\}_{PB} &= &  \varepsilon^{ij}\left[ D_i \alpha^e h_j - D_i\alpha^h e_j+ \mu \alpha^e \pi_i\times h_j + \left(\hat\Lambda \alpha^e- 2m^2\sigma \alpha^\pi\right) e_i\times e_j \right] \nonumber \\
&&+\, \varepsilon^{ij}\left[ \mu\alpha^\pi e_i\times h_j - \left(m^2\alpha^\pi + \mu\alpha^h + 2m^2\sigma \alpha^e\right)\pi_i\times e_j \right] \, ,  \nonumber \\
\left\{\phi(\alpha),\psi_2 \right\}_{PB} &= & \varepsilon^{ij}\left[D_i\alpha^e \pi_j -D_i\alpha^\pi e_j+ \mu\alpha^e \pi_i\times \pi_j  + \alpha^e h_i\times e_j  \right] \\
&& \!\!\!\!\!\!\!\!\!\!\!\!\!\!\!\!\!\!\!\!+\, \varepsilon^{ij}\left[\left(\frac{m^2}{\mu} \alpha^e - \mu\alpha^\pi\right)e_i\times\pi_j  +
\left(\alpha^h + \frac{2m^2\sigma}{\mu} \alpha^e + \frac{m^2}{\mu} \alpha^\pi\right) e_i\times e_j \right]\, ,  \nonumber
\end{eqnarray}
where, as for TMG, 
\be
D_i \alpha^r a_j^s \equiv \partial_i \alpha^r a_j^s - \alpha^r \Omega_i \times a_j^s\, . 
\ee
This again gives us a $14\times 14$ $\bP$-matrix of the general form (\ref{bPform})  with an $11\times 11$ antisymmetric sub-matrix $\bQ$ that is 
of the general form (\ref{bQformNMG}), but with column vectors $v_I$  that we now read off from  (\ref{vsGMG}). 

Once again all dependence on both $\sigma$ and $\Lambda_0$ is contained in those  terms in $v_I$ that are quadratic in canonical variables.  In the absence of these quadratic terms both vectors $v_I$ would be in the column space of $Q$ and the rank of $\bQ$ would then be the same as the rank of $Q$ (i.e. $4$).  However there {\it are}  quadratic terms,  and however we choose $\sigma$ and $\Lambda_0$ the vectors $v_I$ are linearly independent and no linear combination of them  is in the column space of $Q$. The rank of $\bQ$ is therefore $8$,  {\it independently of the values of $\sigma$ or $\Lambda_0$}. As for NMG,  this implies that the dimension of the physical phase space, per space point,  is  $24-14-6=4$. This is consistent with the linearized analysis of the generic GMG model,  and with the Hamiltonian results of  \cite{Blagojevic:2010ir}, 
but it also applies in the $\sigma=\Lambda_0=0$ limit that yields TNMG, and in that case it is not consistent with the linearized analysis of  
\cite{Andringa:2009yc,Dalmazi:2009pm}.  We conclude  that TNMG suffers from a linearization instability. 

\section{Discussion}
\setcounter{equation}{0}

We have shown that  the action for the 3D ``general massive gravity'' (GMG) model incorporating  both ``topologically massive gravity'' (TMG) and ``new massive gravity'' (NMG) \cite{Bergshoeff:2009hq} can be written as the integral of a Lagrangian 3-form constructed from one-form  fields (including a  3-vector dreibein)  and their exterior derivatives. The action  is then defined without the need for a metric, or even a density. 
We should stress that this reformulation of 3D massive gravity models depends on the special combination of curvature-squared invariants that  appear in the NMG/GMG action, which is remarkable because this combination  was not invented for this purpose. It would be interesting to see if a similar metric-independent formalism is possible for 3D massive supergravity \cite{Andringa:2009yc,Bergshoeff:2010mf,Bergshoeff:2010ui}.

We have called these metric-independent actions ``Chern-Simons-like''  because of their similarity to Chern-Simons (CS) theories of gravity. Strictly speaking, they define a generalization of  3D massive gravity models  because equivalence to the usual actions can be established only if the 
dreibein field is assumed to be invertible. This is also  how Chern-Simons (CS) theories of gravity become equivalent to standard metric theories of gravity. The difference, from the perspective of this paper,  is that CS theories require special coefficients for the various terms in the action, with the result that there are no local degrees of freedom. 

The absence of local degrees of freedom in CS gravity models is also apparent from their  Hamiltonian formulation, which can be found directly by a time/space decomposition; the phase space dimension, per space point, is exactly twice the number of local phase-space constraints, all of which are ``first-class'', so the dimension, per space point, of the physical phase space is zero. In contrast, additional constraints are needed for the Hamiltonian formulation of  Chern-Simons-like models.  A single additional, ``secondary'',  constraint suffices for TMG \cite{Carlip:2008qh,Blagojevic:2008bn}, and  this leads to the what we  have called the ``minimal'' Hamiltonian form of this model.  We have used the Chern-Simons-like formulation of NMG and GMG   to  find an analogously ``minimal''  Hamiltonian formulation  requiring two additional constraints. 

Secondary  constraints are needed in the Hamiltonian formulation whenever the field equations imply constraints on the canonical variables that are not already imposed by the time-components of the one-form fields used to construct the Lagrangian. Since these are constraints on the space-components only, imposing them via Lagrange multipliers  leads to an action that is no longer manifestly invariant under 3-space diffeomorphisms in the sense that (in contrast to the CS case) it is not the time/space decomposition of a manifestly covariant Lagrangian. This is not a problem if the new field equations imply the vanishing of the new Lagrange multipliers, because the new field equations are then equivalent to the original ones. For TMG it is easy to show that this is  precisely what happens. In the NMG/GMG case, however,  the new Lagrange multipliers are zero only on one branch
of the solution space of the new equations, with other branches meeting it at ``partially massless'' vacua,  as has also been found in the Hamiltonian approach of \cite{Afshar:2011qw}. Our investigations  led us to conclude that equivalence to NMG/GMG holds only on this one branch.

Fortunately, this ``branch equivalence''  of our Hamiltonian formulation of NMG and GMG is  sufficient for our main  purpose, which is a determination of 
the number of local degrees of freedom of these models. The results of this computation, for all of the models considered in this paper, are summarized in the following table, 
where each model is defined by the combination of invariants that it includes; these are  the Einstein-Cartan (EC) action for 3D GR, the Lorentz-Chern-Simons (LCS) term of 3D conformal gravity and the NMG curvature squared-invariant which, by itself, defines ``massless NMG" (mNMG):
\begin{table}[h]
\begin{center}
\begin{tabular}{ c | c | c || l | c }
\, $L_{EC}$ \, & \, $L_{LCS}$ \, & $L_{mNMG}$ & Name & Degrees of
Freedom \\ \hline
x &  &  & Einstein-Cartan & 0 \\
 & x &  & Conformal Gravity & 0 \\
 x & x &  & TMG & 1 \\
 x &  & x & NMG & 2 \\
  & & x & massless NMG & 2 \\
x & x & x & GMG & 2 \\
 & x & x & TNMG & 2 \\
\end{tabular}
\end{center}
\end{table}

These results confirm those of  \cite{Blagojevic:2010ir,Blagojevic:2011qc,Afshar:2011qw}, in particular the absence of  any discontinuity in the number of local degrees of freedom in the special case in which a linearized analysis yields ``partially massless'' gravitons.  
There are other special cases, however, and we have focused  on the limit of GMG that  yields  ``topologically new  massive gravity'' (TNMG),  where a linearized analysis also exhibits an apparent reduction in the number of local degrees of freedom. Our results show, as expected,  that this is an artefact of the linearized approximation and hence that TNMG suffers from a linearization instability.  

Our result for TNMG also applies to its  parity-preserving ``massless NMG'' limit. This model was  argued in  \cite{Deser:2009hb} to be renormalizable but the argument 
depends on the accidental linearized  invariance  of the linearized theory. This gauge invariance is ``accidental'' because the non-linear theory is certainly not Weyl invariant, although it does  have a ``conformal covariance'' property \cite{Bergshoeff:2009hq} that  explains {\it why} the linearized theory is linearized Weyl-invariant  \cite{Bergshoeff:2010ad}.  Similar considerations explain why linearized TNMG also has an accidental linearized Weyl invariance. 

From this discussion, it should be clear why the number of local degrees of freedom of  interacting massive gravity theories does not change discontinuously in most of the limits 
in which the linearized theory is discontinuous. We should expect discontinuities in decoupling limits: the number of local degrees of freedom is reduced in the TMG limit of GMG, and  the EC limit of TMG,  because we are taking a limit in which one degree of freedom becomes inaccessible; this is not a physical discontinuity. 
The only real discontinuity occurs when the non-linear theory acquires an enhanced gauge invariance, and this occurs {\it only} in limits that yield 
3D conformal gravity. These considerations explain the results of the above table. 

Another result of this paper is  an  alternative Chern-Simons-like  form of the action for TMG and GMG  in which the Lagrange multipliers for the secondary constraints are promoted to new one-form fields. In the TMG case, this action is just the sum of the CS actions for 3D GR and 3D conformal gravity. 
As for our Hamiltonian form of NMG and GMG, there is only a ``branch-equivalence''  to the original CS-like actions, but this just means that we 
have a slightly new 3D massive gravity model that is known to be unitary at least when linearized about one branch of its solutions.  One might think that a time/space decomposition of  this alternative action would lead to a Hamiltonian formulation that  preserves the manifest 3-space covariance
in the same way as CS  models. However,  since the space components of the  new one-form fields are zero as a consequence of the field equations
we now need new non-covariant secondary constraints, leading to a  Hamiltonian formulation that is now ``non-minimal''  but otherwise equivalent to the minimal formulation. 

Following the work of  \cite{Bergshoeff:2009hq}  on 3D massive gravity models,  progress has been made towards the construction of  a non-linear and ghost-free 4D theory of massive  gravity, e.g. \cite{Chamseddine:2010ub} and  \cite{deRham:2010kj}. The model presented in the latter paper has received much  attention although it was, for a while, a challenging technical problem to prove  that it is ghost-free; see e.g.   \cite{Kluson:2012wf} and references therein.  In recent work \cite{Hinterbichler:2012cn}, this 4D massive gravity model has  been  reformulated in a vielbein language that is reminiscent of our CS-like formulation of 3D massive gravity models, so it would be interesting to see  whether the methods that we have used here could also be used to simplify its Hamiltonian formulation. 

\bigskip\bigskip\bigskip

\noindent {\bf Acknowledgements} OH, PKT and BZ are grateful  for the hospitality of  the Isaac Newton Institute for Mathematical Sciences, where they were participants in the program {\it Mathematics and Applications of Branes in String and M-theory} at various periods during the completion of this work.  B.Z. also acknowledges support from 
grant no.11104324 of the  National Natural Science Foundation of China. O.H. is supported by the 
DFG Transregional Collaborative Research Centre grant TRR 33
and the DFG cluster of excellence ``Origin and Structure of the Universe".
\vskip .1truecm

\bigskip
\noindent{\bf Note added}. Following the posting of the original version of this paper to the arXiv,  a revised version of  arXiv:1208.0339 was posted, with a revised title, in which a Hamiltonian analysis of the model referred to here as ``massless NMG'' is presented \cite{Deser:2012ci}, with conclusions that are in accord with 
ours.


\providecommand{\href}[2]{#2}\begingroup\raggedright\endgroup

\end{document}